%% file: document.tex
\documentclass{article}
\usepackage{arxiv}
\pdfoutput=1

\usepackage{amsmath} 
\usepackage{amsfonts} 
\usepackage{bbm} 
\usepackage{dsfont} 
\usepackage{bm}
\usepackage[per-mode=symbol]{siunitx}
\usepackage{multirow}
\usepackage{caption} %

\newcommand{\R}{\mathbb R}
\newcommand{\setY}{\mathcal Y}

\input{libraries.tex}
\input{macros_and_commands.tex}

\begin{document}

\title{CHoKI-based MPC for blood glucose regulation in Artificial Pancreas}

\author{
    Beatrice Sonzogni\\
    Department of Management, Information and Production Engineering\\
    University of Bergamo (24044 Dalmine, Bergamo, Italy)\\
    \texttt{beatrice.sonzogni@unibg.it}\\
    \And
    Jos\'{e} Mar\'{i}a Manzano\\ 
    Department of Engineering\\
    Universidad Loyola Andaluc\'{i}a (41704 Dos Hermanas, Seville, Spain)\\
    \texttt{jmanzano@uloyola.es}\\
    \And
    Marco Polver\\
    Department of Management, Information and Production Engineering\\
    University of Bergamo (24044 Dalmine, Bergamo, Italy)\\
    \texttt{marco.polver@unibg.it}\\
    \And
    Fabio Previdi\\
    Department of Management, Information and Production Engineering\\
    University of Bergamo (Via G. Marconi 5, 24044, Dalmine (BG), Italy)\\
    \texttt{fabio.previdi@unibg.it}\\
    \And
    Antonio Ferramosca\\
    Department of Management, Information and Production Engineering\\
    University of Bergamo (24044 Dalmine, Bergamo, Italy)\\
    \texttt{antonio.ferramosca@unibg.it}\\
}

\date{}

\maketitle

\begin{abstract}
    This work presents a Model Predictive Control (MPC) for the artificial pancreas, which is able to autonomously manage basal insulin injections in type 1 diabetic patients. Specifically, the MPC goal is to maintain the patients' blood glucose level inside the safe range of 70-180 mg/dL, acting on the insulin amount and respecting all the imposed constraints, taking into consideration also the Insulin On Board (IOB), to avoid excess of insulin infusion. MPC uses a model to make predictions of the system behaviour. In this work, due to the complexity of the diabetes disease that complicates the identification of a general physiological model, a data-driven learning method is employed instead. The Componentwise H\"{o}lder Kinky Inference (CHoKI) method is adopted, to have a customized controller for each patient. For the data collection phase and also to test the proposed controller, the virtual patients of the FDA-accepted UVA/Padova simulator are exploited. The proposed MPC is also tested on a modified version of the simulator, that takes into consideration also the variability of the insulin sensitivity. The final results are satisfying since the proposed controller reduces the time in hypoglycemia (which is more dangerous) if compared to the outcome obtained with the standard constant basal insulin therapy provided by the simulator, satisfying also the time in range requirements and avoiding long-term hyperglycemia events.
\end{abstract}


\input{1_introduction.tex}
\input{2_problem_statement.tex}
\input{3_choki_based_robust_mpc.tex}
\input{4_results.tex}
\input{5_conclusion.tex}

\bibliographystyle{plain}
\bibliography{references_books.bib, references_thesis.bib, references_GO.bib, references_BBO.bib, references_PBO.bib, references_nonspecific.bib, references_our_contributions.bib, references_case_study.bib, reference_real_paper.bib}.

\end{document}

%% file: libraries.tex
\usepackage{graphicx}
\usepackage{epstopdf}
\DeclareGraphicsExtensions{.eps,.png,.jpg,.pdf}
\usepackage{float}
\usepackage{subfig} 
\graphicspath{{./img/}{./}}
    
\usepackage{amsmath}
\usepackage{amsthm, thmtools} 
\usepackage{amssymb}
\usepackage{mathtools}
\usepackage{mathrsfs}
\allowdisplaybreaks[1] 

\usepackage[latin1]{inputenc} 
\usepackage{amsfonts}
\usepackage{textcomp}

\usepackage[table]{xcolor}
\usepackage{colortbl} 
\definecolor{LightCyan}{rgb}{0.70,1,1}

\usepackage{multirow}
\usepackage{tabulary}
\usepackage{rotating}

\usepackage{comment}

\usepackage{caption}

\usepackage{algorithm}
\usepackage{algpseudocodex}

\usepackage[shortlabels, inline]{enumitem}

\usepackage{setspace}

\usepackage{hyperref} 

%% file: macros_and_commands.tex
\theoremstyle{thmstyletwo} 

\renewcommand{\min}[1]{\mathchoice
  {\underset{#1}{\operatorname{min}}\,}%
  {\operatorname{min}_{#1}}%
  {\operatorname{min}_{#1}}%
  {\operatorname{min}_{#1}}%
}
\renewcommand{\max}[1]{\mathchoice
  {\underset{#1}{\operatorname{max}}\,}%
  {\operatorname{max}_{#1}}%
  {\operatorname{max}_{#1}}%
  {\operatorname{max}_{#1}}%
}

\newcommand{\blue}[1]{{\textcolor{blue}{#1}}}

\newcounter{algsubstate}
\renewcommand{\thealgsubstate}{\alph{algsubstate}}

\algnewcommand{\IfThenElse}[3]{
  \State \algorithmicif\ #1\ \algorithmicthen\ #2\ \algorithmicelse\ #3}
\algnewcommand{\IfThen}[2]{
  \State \algorithmicif\ #1\ \algorithmicthen\ #2}

\hypersetup{pdfpagemode={UseOutlines},
  bookmarksopen=true,
  bookmarksopenlevel=0,
  hypertexnames=false,
  colorlinks=true,
  citecolor=blue,
  linkcolor=blue,
  urlcolor=black,
  pdfstartview={FitV},
  unicode,
  breaklinks=true,
}

%% file: 1_introduction.tex
\section{Introduction}
\label{sec:Introduction}
Diabetes is a common chronic metabolic disorder characterized by the body's inability to correctly balance the Blood Glucose (BG) level, due to the absence or not enough insulin production by the pancreas. In particular, Type 1 Diabetes (T1D) is characterized by the autoimmune destruction of the insulin-producing beta cells. This drives the patient into a state of hyperglycemia, so when the BG level is above 180~\si{\milli\gram / \deci\liter}, which has long-term complications, such as cardiovascular disease, neuropathies or kidney damage.
Thus, T1D patients require daily insulin injections to maintain the BG level inside the euglycemic range (i.e. between 70 and 180~\si{\milli\gram / \deci\liter}). Below this threshold, the patient is in a state of hypoglycemia, which is dangerous in the short-term since can even lead to the diabetic coma~\cite{diabetes}.
In order to ease patients' and caregivers' life, the therapy is trying to get more and more automatised, miming the functioning of a healthy pancreas. Specifically, the Artificial Pancreas (AP) implements such a treatment in a closed loop. It is a system made of three components: the sensor that measures the glucose at the interstitial level every few minutes (Continuous Glucose Monitoring, CGM), and the control algorithm that computes the insulin amounts that are then injected into the subcutaneous tissue through the pump.
The APs currently on the market are hybrid closed loop systems, since the administration of the basal insulin (which is the small, continuous and constant amount injected to manage the BG in fasting periods) is automatic, while for postprandial boluses (i.e. bigger amount injected at meal times to face the BG increase due to carbohydrate ingestion or when the BG level is unexpectedly too high) it still requires the manual intervention of the patients~\cite{moon}.

Model Predictive Control (MPC), due to its predictive capability and the possibility to add constraints to the problem, is one of the most widely used control algorithms for the AP.
MPC is a control strategy that uses a dynamic model to forecast the future behavior of a system. Based on this prediction, it calculates the best sequence of control actions at every sampling time, by solving a finite horizon optimal control problem. Then, only the first value of the control action sequence is applied to the system and the process is repeated at each sampling time, in a receding horizon fashion~\cite{rawlings}.
Over the last few years, the use of MPC as a control algorithm for AP has been extensively studied and tested~\cite{del2019deployment,toffanin2013artificial,hovorka2004nonlinear,abuin2020artificial,gondhalekar2016periodic,gonzalez2020stable,shi2018adaptive,hajizadeh2019plasma,gonzalez2017impulsive,soru2012mpc}, thanks to its capability to anticipate unwanted fluctuations in glycemic levels and to calculate the amount of insulin to be injected, taking into account all the constraints. 

T1D is a disease that varies both among and within patients and this is due to differences in blood glucose response to meals or insulin, which can also vary according to daily state. Identifying a general model to describe the insulin-glucose system is therefore difficult. This work aims to use data-driven approaches, that is, to use the current and past data of a patient to obtain the future BG, and then to be able to calculate the correct amount of basal insulin to be delivered. This facilitates and improves T1D management by providing a customised MPC algorithm for the AP. Recently, different types of learning-based MPCs have been proposed in the literature~\cite{hewing2020learning}, which are based on different learning methods.
Specifically, we use the Componentwise H\"{o}lder Kinky Inference (CHoKI) method, a nonparametric learning technique that favors the design of robust MPCs that are stable by design~\cite{manzano}.

In this case, starting from the work proposed in~\cite{ifac}, in the MPC optimization problem is considered also a dynamic safety constraint on the maximum basal insulin value, which is based on the Insulin On Board (IOB). To take into account the insulin amount of the boluses that is still active in the patient, when computing the quantity of the basal corrections, to reduce the risk of hypoglycemic events.

The virtual patients of the UVA/Padova simulator~\cite{DMMSR} are exploited to collect the data needed to learn the system behaviour and also to test the proposed control algorithms. This is a simulator accepted by the Food and Drug Administration (FDA) as a substitute for pre-clinical studies and it contains populations of virtual subjects; in particular, we have used the adults with T1D.

The same proposed CHoKI-based MPC has been also tested on patients whose insulin sensitivity varies intra-subjects throughout the day.

The rest of this work is structured as follows: Section~\ref{sec:sec:problem_statement} presents the CHoKI learning method and the version tailored to the T1D patient case. Section~\ref{sec:choki_based_robust_mpc} analyses the proposed model predictive control problem. The implementation of the designed controller in the UVA/Padova simulator is presented in Section~\ref{sec:results}, and Section~\ref{sec:conclusion} concludes the paper. 
\\

\noindent
\emph{Notation} 

A set of integers $[a,b]$ is denoted $\mathbb{I}^b_a$, $\mathbb{R}^n$ is the set of real vectors of dimension $n$ and $\mathbb{R}^{n \times m}$ is the set of real matrices of dimension $n \times m$. 
Given \mbox{$v,w\in\mathbb R^{n_v}$}, the notation $(v,w)$ implies $[v^T,w^T]^T$ and $v\leq w$ implies that the inequality holds for every component. $\|v\|$ stands for the Euclidean norm of $v$ and $|v|=\{w:w_i=|v_i|,\forall i\}$. Given two sets $A,B$, $A\ominus B$ denotes the Pontryagin difference. Their Cartesian product is denoted $A \times B = \{(x,y)|x\in A, y \in B\}$.
The positive box $\mathds{B}(v) \subset \mathbb{R}^{n_v}$ of radius $v$ is defined as~$\mathds{B}(v)=\{y: 0 \leq y \leq v\}$ and the ball $\mathcal{B}(v) \subseteq \mathbb{R}^{n_v}$ of radius $v$ is defined as~$\mathcal{B}(v)=\{y: |y| \leq v\}$.
An $n,m$-dimensional matrix of ones is denoted $\mathds{1}_{n \times m}$. The $i$th row of a matrix $M$ is denoted $M_i$. 

%% file: 2_problem_statement.tex
\section{Problem statement}
\label{sec:sec:problem_statement}

The analysed problem is based on the CHoKI formulation proposed in \cite{ifac}, which is briefly reported in the following.

The system is a sampled continuous-time one, described by an a priori unknown discrete-time model, whose measured output is $y(k) \in \mathbb{R}^{n_y}$ and whose input is $u(k) \in \mathbb{R}^{n_u}$. In this case, there is one output ($n_y=1$) which is the glucose level, in \si{\milli\gram / \deci\liter}, and there are two inputs ($n_u=2$), which are the meal ($u_1$, the not controllable one) in \si{\gram} of carbohydrates and the insulin ($u_2$, the controllable one) in \si{\pico\mol}. A sampling time of 5 minutes is considered.

The measured output can be modeled as a nonlinear autoregressive exogenous (NARX) model, with the following state-space representation: 
\begin{equation} \nonumber
    y(k+1)=f(x(k),u_1(k),u_2(k)) + e(k),
\end{equation} 
where $e(k) \in \mathbb{R}^{n_y}$ is process noise and the regression state $x \in \mathbb{R}^{n_x}$ is
\begin{equation}\label{eq:state}
    x(k)=\Big(y(k), \dots, y(k-n_a), u_1(k-1), \dots, u_1(k-n_b),    u_2(k-1), \dots, u_2(k-n_c)\Big),
\end{equation}
where $n_a\in \mathbb{N}_0$ is the memory horizon for the glucose values,~$n_b\in \mathbb{N}_0$ for the meals and~$n_c\in \mathbb{N}_0$ for the basal insulin injections. 
The arguments of $f$ are then aggregated into  $w=(x,u_1,u_2) \in \mathbb{R}^{n_w}$ so that it is possible to build a data set of~$N_\mathcal{D}$ observations, denoted~$\mathcal{D}=\{(y_{k+1},w_k)\}$, for $k=1, \ldots, N_{\mathcal{D}}-1$.

\subsection{Componentwise H\"{o}lder Kinky Inference (CHoKI)} \label{sec:choki}
The purpose of this subsection is to be an introduction to the choice of learning method.
Kinky Inference (KI) \cite{manzano_lacki} is a class of learning approaches that includes Lipschitz interpolation, which is a technique based on Lipschitz continuity of the ground truth function.
There exists an extension of the Lipschitz continuity, named H\"{o}lder continuity, defined as follows: 

\textbf{Definition 1} (H\"{o}lder continuity)\textbf{.} \textit{A function $f:\mathcal{W} \rightarrow \mathcal{Y}$ is H\"{o}lder continuous if there exist two real constants $L \geq 0$ and $0 < p \leq 1 $ such that for all $w_1, w_2 \in \mathcal{W}$,}
\begin{equation} 
    \|f(w_1)-f(w_2)\|\leq L \|w_1-w_2\|^p ,
\end{equation}
\textit{where $L$ represents the smallest Lipschitz constant and $p$ is called the H\"{o}lder exponent, $\mathcal{W}\subseteq \mathbb{R}^{n_w}$ is the input space and $\mathcal{Y} \subseteq \mathbb{R}^{n_y}$ is the output space. \\
In the case of $p=1$, it means to have Lipschitz continuity}~\cite{manzano}.

To catch different variations of the output according to the changes of each component of the input regressor, the Componentwise H\"{o}lder Kinky Inference (CHoKI)~\cite{manzano} can be implemented. This method is based on the componentwise H\"{o}lder continuity, which considers matrices $\mathcal{L}$ and $\mathcal{P}$, instead of the H\"{o}lder constant $L$ and exponent $p$. 
This is useful in cases where a function may have sudden variations along one dimension of the input, while changing smoothly along other input dimensions. 

\textbf{Definition 2} (Componentwise H\"{o}lder continuity)\textbf{.} \textit{Given the matrices $\mathcal{L}$ and $\mathcal{P}\in\R^{n_y\times n_w}$, a function $f:\mathcal{W} \rightarrow \mathcal{Y}$ is componentwise $\mathcal{L}$-$\mathcal{P}$-H\"{o}lder continuous if $\forall w_1,w_2 \in \mathcal{W}$ and $\forall i \in \mathbb{I}_1^{n_y}$}
%
\begin{equation} \label{eq:holder_continuous}
    |f(w_1)-f(w_2)| \leq \mathfrak{d}_\mathcal{L}^\mathcal{P} (|w_1-w_2|)
\end{equation}
\textit{where }
\begin{equation}
    \mathfrak{d}_\mathcal{L}^\mathcal{P} (w):= \big(a:a_i=\sum_{j=1}^{n_w} \mathcal{L}_{i,j} w_j^{\mathcal{P}_{i,j}},\forall i \in \mathbb{I}_1^{n_y}\big) .
\end{equation}

Then, assuming that $f$ is H\"{o}lder continuous and given a data set $\mathcal D$ of inputs/outputs observations, the CHoKI predictor for a query~$q\in\mathbb R^{n_w}$ is: 
\begin{equation}  \label{eq:predictor}
    \hat{f}(q;\Theta,\mathcal{D})= {\frac{1}{2}} \min{i=1,\dots,N_\mathcal{D}} \Big(\tilde{y}_i+\mathfrak{d}_\mathcal{L}^\mathcal{P} \big(|q-w_i|\big)\Big) + {\frac{1}{2}} \max{i=1,\dots,N_\mathcal{D}} \Big(\tilde{y}_i-\mathfrak{d}_\mathcal{L}^\mathcal{P} \big(|q-w_i|\big)\Big) ,
\end{equation}

where $\Theta=\{\mathcal{L},\mathcal{P}\}$ and $\hat{f}$ is still componentwise $\mathcal{L}$-$\mathcal{P}$-H\"{o}lder continuous.

According to~\eqref{eq:predictor} it is possible to predict a new output $\hat{y}(k+1)=\hat{f}\big(w(k);\Theta,\mathcal{D}\big)$. 
Then, the prediction model can be formulated in state-space as follows: 
\begin{equation} \label{eq:state_space_pred}     
    \begin{array}{ll}
    \hat{x}(k+1)=\hat{F}\big(x(k),u_1(k),u_2(k)\big) \\
    \hat{y}(k)=M\hat{x}(k) 
    \end{array}
\end{equation} 
where $\hat{F}\big(x(k),u_1(k),u_2(k)\big)=\big(\hat{f}(x(k),u_1(k),u_2(k)),y(k),\ldots,y(k-n_a+1),u_1(k),\ldots,u_1(k-n_b+1),u_2(k),\ldots,\\u_2(k-n_c+1)\big)$ and $M=[I_{n_y},0,\ldots,0]$. 

If the matrices $\mathcal{L}$ and $\mathcal{P}$ are unknown a priori, they must be estimated. To do that, an optimization problem is solved offline, splitting the data set~$\mathcal D$ into two disjoint data sets: $\mathcal{D}_\mathrm{train}$ for the estimation and $\mathcal{D}_\mathrm{test}$ for the validation. The optimization problem is:  
\begin{subequations}    
\label{eq:optimizazion_problem_choki}
\begin{align}  
   \Theta&=\arg \min{\Theta} g(\Theta,\mathcal{D}_\mathrm{train},\mathcal{D}_{\mathrm{test}}) \\
\textrm{s.t. } & |\tilde{y}_i-\tilde{y}_j| \leq \mathfrak{d}_\mathcal{L}^\mathcal{P} (|w_i-w_j|), \quad \forall w_i, w_j \in \mathcal{W_D}, \, w_i\neq w_j  \\
 & 0< \mathcal{P}_{ij}\leq 1, \, \mathcal{L}_{ij}>0, \quad i \in \mathbb{I}_1^{n_y}, \, j \in \mathbb{I}_1^{n_w},
\end{align}
\end{subequations}
where~$\mathcal W_\mathcal D$ represents the input data points in~$\mathcal D$. The cost function $g(\Theta,\mathcal{D}_\mathrm{train},\mathcal{D}_\mathrm{test})$ to be minimized is: 
\begin{equation} 
    g(\Theta,\mathcal{D}_\mathrm{train},\mathcal{D}_\mathrm{test})= {\frac{1}{N_{\mathcal{D}_\mathrm{test}}}} \sum_{i=1}^{N_{\mathcal D_\mathrm{test}}} \| \hat{f}(w_i;\Theta,\mathcal{D}_\mathrm{train}) - \tilde{y}_i \|^2 ,
\end{equation}
being $\hat{f}(w_i;\Theta,\mathcal{D}_\mathrm{train})$ the predictions made with the CHoKI in~\eqref{eq:predictor} (computed with the data in~$\mathcal{D}_\mathrm{train}$), which are compared to $\tilde{y}_i$, that are the measured values of the noisy data set~$\mathcal{D}_\mathrm{test}$.

\subsection{CHoKI implementation for T1D patient}
In this subsection, the CHoKI method explained in Section~\ref{sec:choki} is designed to learn the dynamics of the T1D patient. 

To this aim, to implement the CHoKI strategy, the first step is the data collection.
This is a fundamental phase since the quality of the generated data set will affect the performance of the CHoKI predictor and thus the functioning of the controller. 
This is done by employing the virtual T1D adult patients of the UVA/Padova simulator. For each of them, several simulations were made, varying the initial glycemic condition, the basal insulin quantity (from 0 to~\SI{500}{\pico\mol}) and the carbohydrates of the meals (with the post-prandial boluses, given~\SI{20}{\minute} after the meal time). All these simulations were set to obtain an appropriate distribution of the points in the space, looking at the input-output representation.
The simulator allows the inclusion of some noises on the sensor and on the pump, to perform more realistic simulations. Specifically, the available virtual typical commercial CGM was selected as a sensor, with auto-regressive noise with inverse Johnson transform distribution. 
The noise on the virtual pump is normally distributed with a mean of~\SI{0}{\pico\mol} and a standard deviation of 0.1.  
Also, an error with a normal distribution with a standard deviation equal to 30\% of the meal amount is added to the carbohydrate estimation.

Only the relationship between BG, meals and basal insulin is considered, as the aim of the proposed controller is to manage basal insulin injections automatically, while meal boluses are delivered manually (assuming they are a function of meals).
The CHoKI requires the data to be in the right NARX shape, thus the model orders $n_a, n_b$ and $n_c$ have to be identified and this is done through a cross-validation procedure.

In particular, many combinations of model orders were tested. The Mean Squared Error (MSE) among the 1-step ahead predictions and the actual values were measured on an unused 21-day data set. 
The selected orders were chosen based on the lowest MSEs, but a trade-off with model complexity was also considered to avoid the risk of overfitting. 
The resultant orders are $n_a=5$, $n_b=9$ and $n_c=3$, being each sampling time~\SI{5}{\minute} long.


To obtain the predictions employing~\eqref{eq:predictor}, the hyperparameters $\Theta=\{\mathcal{L},\mathcal{P}\}$ must be estimated according to~\eqref{eq:optimizazion_problem_choki}.
We assumed to have $\mathcal{P}= \mathds{1}_{n_y \times n_w}$ and thus the optimization problem is solved to obtain just the values of the matrix $\mathcal{L}$. In this case, only three values are estimated:~$L_a\in\R$ for the glucose part, ~$L_b\in\R$ for the meals and~$L_c\in\R$ for the insulin. 
Therefore, $\mathcal{L}$ contains those three values repeated to reach the right dimension (i.e. the regressor length $n_w$), thus~$\mathcal{L} = [L_a \mathds{1}_{n_a}; L_b  \mathds{1}_{n_b}; L_c  \mathds{1}_{n_c}]$.

Some a priori knowledge was utilized in order to set the constraints of the optimization problem: as the~$\mathcal{L}$ initial value the Lipschitz constant~$L$ was exploited, which is obtained from the LACKI (Lazily Adapted Constant Kinky Inference) method~\cite{manzano_lacki}, based on the H\"{o}lder continuity property. The upper and lower bounds for $L_a$, $L_b$ and $L_c$ were set as~[10;10;10] and~[0;0.9;0.09], respectively, thanks to previous analyses.

The \texttt{fmincon} MATLAB function was implemented to solve the optimization problem~\eqref{eq:optimizazion_problem_choki}.
For each patient, once the~$\mathcal{L}$ is found, the model is validated on a new data set, to verify its ability to predict future BG values, comparing them with the real outputs.
For each subject, the resulting $\mathcal{L}$, the $u_\mathrm{ref}$ and the $L$ are reported in Table~\ref{tb:mpc_settings}. 

\begin{figure}
\begin{center}
\includegraphics[width=\textwidth]{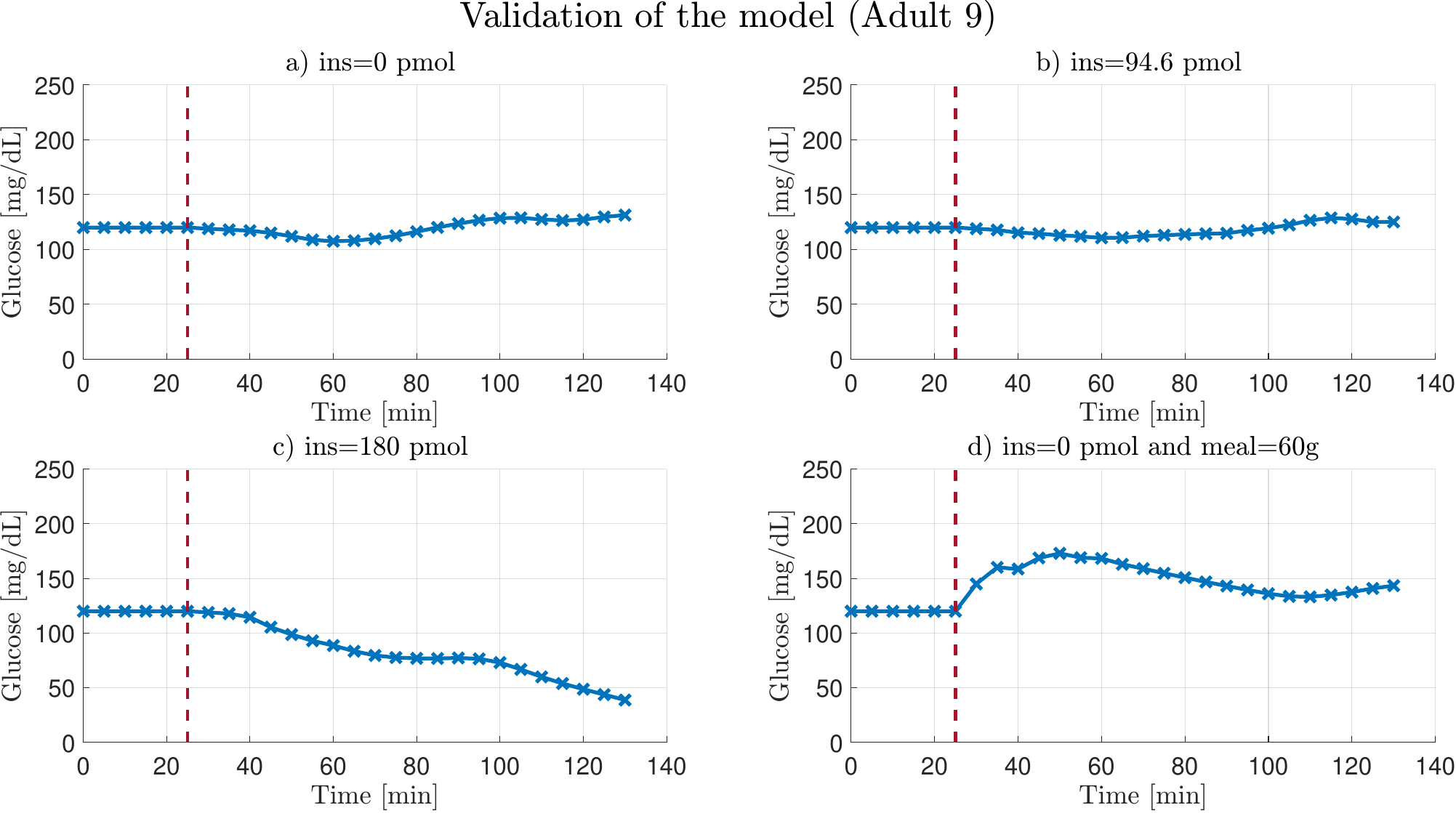}
\caption{In each graph, the vertical dashed red line marks the end of the fixed regressor, when the inputs displayed in the titles are applied. The blue line is the glucose trend. In a) the glucose increases a bit, due to the absence of basal insulin. In b) the glucose remains stable since the insulin amount is the reference value and equal to the regressor values. In c) the glucose decreases due to the basal amount of 180 \si{\pico\mol}. In d) the glucose increases due to the presence of the meal and no basal insulin. }
\label{fig:proveModello}
\end{center}
\end{figure}

Further analysis was also carried out, starting with a fixed regressor and varying the input values, to ensure that the CHoKI strategy had correctly learned the effect of each input on the output. The fixed regressor has BG set to 120~\si{\milli\gram / \deci\liter}, no meals and constant basal insulin equal to the reference value $u_\mathrm{ref}$ of each patient, obtained from the standard therapy provided by the simulator.
As an example, the results obtained from subject Adult 9 are displayed in Figure~\ref{fig:proveModello}: it can be seen how the glucose trend (blue lines) decreases when the amount of the basal insulin injections increases and it rises when there is carbohydrate ingestion. This holds for all the considered virtual patients.

\begin{table*}[t] 
\begin{center}
\caption{MPC settings}\label{tb:mpc_settings}
\begin{tabular}{|c|c|c|c|c|c|c|c|
c|}
\hline
Adult & \begin{tabular}[c]{@{}c@{}}$u_\mathrm{ref}$ \\ (\si{\pico\mol})\end{tabular} & $N_{\mathcal{D}}$ & \begin{tabular}[c]{@{}c@{}}$L$\\ (LACKI)\end{tabular} & \begin{tabular}[c]{@{}c@{}}$[L_a; L_b; L_c]$\\ (CHoKI)\end{tabular}   & \begin{tabular}[c]{@{}c@{}}$\mu~(90\%)$\\ (\si{\milli\gram\per\deci\liter}) \end{tabular} & $N_c$ & $\epsilon$ & Q \\\hline
\#1 & 122.38 & 4775 & 3.46 & [0.74; 5.46; 0.29]  & 14.83 & 2 & 10 & 1 \\
\#2 & 134.89 & 4950 & 3.28 & [4.89; 3.96; 0.09] & 10.19  & 2 & 20 & 1  \\
\#3 & 149.97 & 4990 & 3.08 & [0.71; 5.45; 0.09]  & 9.29  & 3 & 10 & 1 \\
\#5 & 91.83 & 4156 & 6.56 & [0.84; 5.52; 0.44]  & 13.91  & 2 & 5 & 1 \\
\#6 & 190.22 & 5339 & 3.41 & [4.72; 3.52; 0.09] & 11.27 & 1 & 1 & 1 \\
\#8 & 105.83 & 4703 & 2.58 & [1.08; 5.84; 0.09]  & 7.8 & 3 & 1 & 100  \\
\#9 & 94.59 & 3976 & 3.72 & [1.13; 4.09; 0.09]  & 11.63  & 2 & 1 & 100 \\
\#10 & 124.86 & 4966 & 3.29 & [3; 2; 0.09]  & 10.1  & 1 & 20 & 1 \\
\hline
\end{tabular}
\end{center}
\end{table*}

%% file: 3_choki_based_robust_mpc.tex
\section{CHoKI-based robust MPC}
\label{sec:choki_based_robust_mpc}

The control objective is to drive and maintain the BG level inside the euglycemic zone, which is given by~70 and~$\SI{180}{\milli\gram\per\deci\liter}$, satisfying all the inputs and output constraints. 
The basal insulin amount~$u_2(k)$ must be inside the range~$\mathcal U=\{u:0\leq u\leq \SI{500}{\pico\mol}\},\forall k$. It is the control action, whose values are calculated so that the BG level $y(k)$ remains in the set~$\setY=\{y:55\leq  y \leq \SI{300}{\milli\gram\per\deci\liter}\},\forall k$, not to arrive to extreme hyper- or hypoglycemic conditions.
An additional constraint is set on the maximum value of the basal insulin, according to the IOB estimation.

In this case, the open-loop predictions of the MPC control problem are computed with the CHoKI predictor~\eqref{eq:predictor}, assuming that a physiological model for T1D patients is not available.
To ensure the robustness of the MPC to possible model-system mismatches, the output constraints are tightened at each step according to an error that represents the uncertainty of the predictions based on the data.
The system in closed loop is shown to be Input-to-State Stable (ISS)~\cite[Theorem 3]{manzano} with the proposed controller.

The set of restricted output constraints is given by
\begin{equation} \label{eq:constraints}
    \mathcal{Y}_j = \mathcal{Y}_{j-1} \ominus \mathcal{R}_j,
\end{equation}
along the prediction horizon, $j=1,...,N$.~$\mathcal{R}_j$ are the reachability sets that account for the possible errors in the nominal predictions and $\mathcal{Y}_0=\mathcal{Y}$. 
To compute $\mathcal{R}_j$, the starting point is to consider a sequence of future inputs $u(k+1)$ and $c_1\in \mathbb{R}^{n_y}$, such that 
\begin{equation}
    |y(k+1)-\hat{y}(1|k)| \leq c_1 .
\end{equation}
The difference between a prediction made at time step $k+j$ based on the measurement at step $k$, and the prediction made at step $k$ based on the measurement at step $k+1$, for a given sequence of control inputs, is bounded by the sets
\begin{subequations} 
\begin{equation} 
    |\hat{y}(j|k)-\hat{y}(j-1|k+1)| \in \mathcal{M}_j \subseteq \mathbb{R}^{n_y}  ,
\end{equation}
\begin{equation} 
    |\hat{w}(j|k)-\hat{w}(j-1|k+1)| \in \mathcal{G}_j \subseteq \mathbb{R}^{n_w}  .
\end{equation}
\end{subequations}
The sets $\mathcal{M}$ and $\mathcal{G}$ can be calculated from the equations 
\begin{subequations} 
\begin{equation} 
    \mathcal{M}_j = \mathds{B}\big(\mathfrak{d}^{\mathcal{P}}_{\mathcal{L}}(\mathcal{G}_{j-1})\big)  ,
\end{equation}
\begin{equation} 
    \mathcal{G}_j = \mathcal{M}_j \times \dots \times \mathcal{M}_{\sigma(j)} \times \{0\} \times \dots \times \{0\} ,
\end{equation}
\end{subequations}
with $\sigma(j)=\max(1,j-n_a)$ and $\mathcal{M}_1 = \mathds{B}(c_1)$.
The set $\mathcal{R}_j$ is defined as $\mathcal{R}_j = \{y: |y| \in \mathcal{M}_j \}$ for all $j \in \mathbb{I}^N_1$.

In \cite{manzano} is also shown that $c_j \in \mathbb{R}^{n_y}$ and $d_j \in \mathbb{R}^{n_w}$ are such that~$\mathcal{M}_j = \mathds{B}(c_j)$ and $\mathcal{G}_j = \mathds{B}(d_j)$. Then, the sets $\mathcal{M}_j$ and $\mathcal{G}_j$ can be computed using the recursion 
\begin{subequations} 
\begin{equation} 
    c_j=\mathfrak{d}^{\mathcal{P}}_{\mathcal{L}}(d_{j-1})  ,
\end{equation}
\begin{equation} 
    d_j=(c_j,\dots, c_{\sigma(j)},0,\dots,0) ,
\end{equation}
\end{subequations}
with~$c_1=\mu$ (where $\mu$ is the maximum absolute error obtained in the validation phase) and then,~$\mathcal{R}_j=\mathcal{B}(c_j)$. 

In our specific control problem, an \textit{a posteriori} analysis showed that extreme deviations from nominal predictions are highly unlikely. Then, the value representing the $90^{\mathrm{th}}$ percentile of the probability distribution is used as $\mu$ instead of the maximum error (see Table~\ref{tb:mpc_settings}). 

\textbf{\textit{Remark 1:}} To deal with the infeasibility of possible solutions outside the 90\% region, some slack variables $\bm{\delta}=\{\delta_{min},\delta_{max}\}$ are added to the optimization problem; therefore the constraints on the glucose become $\hat{y}(j|k) \in \mathcal{Y}_{j,\bm\delta},~\forall j\in\mathbb I_1^N$, with 
\begin{equation} \label{eq:constraints_slack}
    \mathcal{Y}_{j,\bm{\delta}}= \{ y: y_{min}(j)-\delta_{min}(j) \leq y \leq y_{max}(j)+\delta_{max}(j)\} ,
\end{equation}
where $y_{min}$ and $y_{max}$ are the extreme values of~$\setY_j$ from~\eqref{eq:constraints}.

\subsection{Terminal ingredients computation} \label{sec:terminal_ingredients}
The tightened constraints are computed as described in the previous section, for all subjects, only once and offline. 
This tightening implies the definition of the length of the control horizon $N_c$, which may vary for each virtual patient.
The control horizon is calculated as the maximum possible value that makes it possible to have a set of constraints that is not empty, but it also takes into account the need to have reasonable ranges according to the system.

A prediction horizon $N_p$ longer than the control horizon $N_c$ is considered to increase the domain of attraction and the predictive ability of the controller, thus~$N_p>N_c$. A local control law for the predictions from $N_c$ to $N_p$ must be established to apply this approach. We use the following:
\begin{equation} \label{eq:uk}
	u=K(\overline{x}-x)+\overline{u} ,
\end{equation}
with $u=(u_1,u_2)$ and where $K\in\mathbb R^{n_u\times n_x}$ is the control gain of a Linear Quadratic Regulator (LQR) and $(\overline{x},\overline{u})$ is an equilibrium point around which the system $\hat F(x,u)$ is linearized. In particular,  $\overline{x}$ is constructed as per~\eqref{eq:state}, using $\overline y=\SI{120}{\milli\gram \per \deci\liter}$ of glucose, and $\overline{u}=(0,u_\mathrm{ref})$. Matrices $A \in \mathbb{R}^{n_x \times n_x}$ and $B \in \mathbb{R}^{n_x \times n_u}$ of the linearized model  $x(k+1)=Ax(k)+Bu(k)$,  are calculated numerically from the input-output data using the CHoKI model. In this way, each element $A(j,i)$ and $B(j,i)$ is obtained by considering that
\begin{equation} \nonumber
   A(j,i) = {\frac{\partial \hat{F}_j}{\partial x_i}}={\frac{\hat{F}_j(\overline{x_i} + \epsilon) - \hat{F}_j(\overline{x_i} - \epsilon)}{2 \epsilon}} , \quad B(j,i) = {\frac{\partial \hat{F}_j}{\partial u_i}} ,
\end{equation}
where $\epsilon$ can be different for each subject (see Table~\ref{tb:mpc_settings}). 
Note that $A(1,1)={\frac{\partial y_{k+1}}{\partial y_k}}$ and $B(1,1)={\frac{\partial y_{k+1}}{\partial u_{1,k}}}$.

\subsection{Insulin On Board}
The MPC algorithm also includes a dynamic safety constraint on the maximum basal insulin value, which is based on the amount of IOB.
The IOB represents the quantity of injected insulin still active in the body, which depends on patient dynamics and on the duration of insulin action (DIA).
The IOB at each sampling time~$k$ can be estimated considering the residuals of the past insulin administration, which means having:
\begin{equation} \label{eq:iob_estimation}
    IOB(k) = a(k-1)u_b(k-1)+ \ldots + a(k-n_{IOB}) u_b(k-n_{IOB}) , 
\end{equation}
where the insulin action curve is represented by~$a$ and the vector of the insulin boluses administration history is~$u_b$. The time of insulin action is~$n_{IOB}$, considered with a sampling time of~$5~\si{\minute}$. Specifically, in this case, it is~$n_{IOB}=72$, which means considering $6~\si{\hour}$ (note that taking into account an insulin duration of $6~\si{\hour}$ is a less conservative approach than considering a duration of $8~\si{\hour}$)~\cite{leon2013postprandial}.

\begin{figure}
    \centering
    \includegraphics[width=0.8\linewidth]{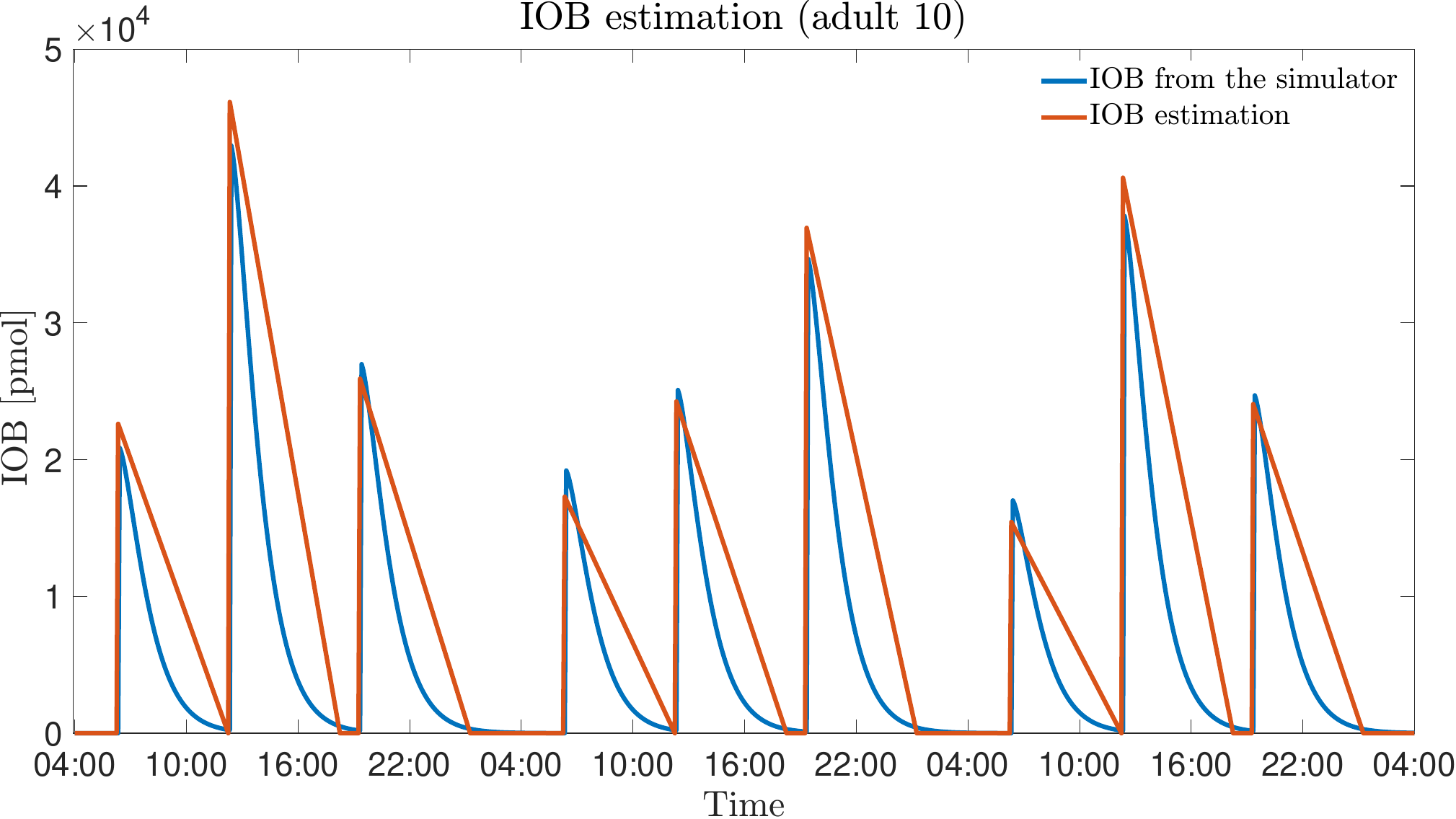}
    \caption{The estimated boluses IOB is represented as the orange line and the IOB computed by the simulator is in blue. This is an example of the virtual patient Adult 10.}
    \label{fig:IOB_estimation_1}
\end{figure}

The upper constraint is considered to limit the basal corrections, in order to avoid giving too much insulin, and thus to reduce the risk of hypoglycemic events. 
This means that the basal insulin amount $u_2$ must be inside the new range
\begin{equation} \label{eq:constraint_u2}
    \mathcal U_2=\{u:0\leq u\leq u_2^\mathrm{max} \} ,
\end{equation}
where $u_2^\mathrm{max}$ is the value of the upper constraint for the basal computation and it comes from:

\begin{equation}
u_2^\mathrm{max}(k,j) = \left\lbrace
\begin{array}{ll}
u_2^\mathrm{lim}-IOB(k,j) & \mathrm{if}\quad u_2^\mathrm{lim}>IOB(k,j)\\
u_\mathrm{ref}&\mathrm{otherwise} 
\end{array}
\right.
\end{equation}
\noindent
where~$u_2^\mathrm{lim}=\SI{500}{\pico\mol}$ is the basal insulin maximum amount that can be injected, and considering the sampling time~$k$ and~$j=0,\dots,N_p-1$~\cite{ellingsen2009safety}. 
The IOB varies along the prediction horizon (i.e. with $j$), which means that the estimations decrease according to the insulin action curve, without considering possible new meal boluses.
This implies that the basal upper bound is equal to the reference value $u_\mathrm{ref}$ when the maximum limit $u_2^\mathrm{lim}$ is less than the insulin still active from the previous boluses injection (i.e. $IOB$), otherwise, the upper bound is equal to the difference between $u_2^\mathrm{lim}$ and $IOB(k)$. 
The weights of the insulin action curve $a$ are obtained from~$\big((DIA-t_b)/DIA\big)$, where~$DIA=6~\si{\hour}$ and $t_b$ is the time passed from the previous bolus. 
In Figure~\ref{fig:IOB_estimation_1} an example of the IOB estimation for virtual Adult 10 is reported. This shows that~\eqref{eq:iob_estimation} approximates quite well the real values (blue line) and the choice of DIA equal to $6~\si{\hour}$ is appropriate. 
The initial IOB value after a bolus could be different between the two curves. This is because the estimated IOB is based on the value of the bolus calculated for the meal, on the other hand, the real IOB is based on the value that is actually injected, which may vary from the calculated value due to pump noise.

\textbf{\textit{Remark 2:}} To address any potential infeasibilities during the MPC resolution, $N_p$ slack variables $\delta_u$ were included in the optimization problem. These were added to the upper bound $u_2^\mathrm{max}$ in equation~\eqref{eq:constraint_u2}, obtaining
\begin{equation} \label{eq:constraints_u2_slack}
    \mathcal U_2'=\{u(j):0\leq u(j)\leq u_2^\mathrm{max}(j)+\delta_u(j) \} , \, \, \forall j\in\mathbb I_0^{N_p-1} .
\end{equation}

\subsection{CHoKI-based MPC implementation}

Starting from the proposal in~\cite{ifac}, the MPC optimization problem which considers also the IOB is set as follows:
\begin{subequations}  \label{eq:mpc}
\begin{eqnarray}
   &&\min{u_2,y_a,\delta_\mathrm{hyper}, \delta_\mathrm{hypo},\delta_\mathrm{min},\delta_\mathrm{max},\delta_u} V_N(\hat{x},u; \Theta, \mathcal{D}) \\
\textrm{s.t. } && \hat{x}(0|k)=x(k) \\
&& \hat{x}(j+1|k) = \hat{F}(\hat{x}(j|k),u_1(j),u_2(j)), \, \, j\in\mathbb I_{0}^{N_c-1} \\ 
&& \hat{x}(j+1|k) = \hat{F}(\hat{x}(j|k),K(\bar x - x(j))+\bar u), \, \,j\in\mathbb I_{N_c}^{N_p-1} \\
&& \hat{y}(j|k) = M \hat{x}(j|k), \, \, j\in\mathbb I_{0}^{N_p-1}\\ 
&& u_2(j) \in \mathcal{U}_2', \, \, j\in\mathbb I_{0}^{N_p-1}, \label{eq:MPC_iob} \\ 
&&\hat{y}(j|k) \in \mathcal{Y}_{j,\bm\delta}, \, \, j\in\mathbb I_{0}^{N_c-1}\\
&&\hat y(j|k) \in \mathcal Y_{N_c,\bm\delta},\, \, j\in\mathbb I_{N_c}^{N_p-1}\\
&& u_1(j)=0,\; j \in \mathbb{I}^{N_p-1}_1\label{eq:nomeal}\\
&& 70 - \delta_\mathrm{hypo} \leq y_a \leq 140 + \delta_\mathrm{hyper} \\
&& \delta_\mathrm{hyper} \geq 0,\; \delta_\mathrm{hypo} \geq 0 \\ 
&& \delta_\mathrm{min}(j) \geq 0,\; \delta_\mathrm{max}(j) \geq 0, \, \, j\in\mathbb I_{0}^{N_p-1} \\
&& \delta_u(j) \geq 0, \, \, j\in\mathbb I_{0}^{N_p-1}
\end{eqnarray}
\end{subequations}
where~\eqref{eq:nomeal} is used since the meals are not predictable,~$\mathcal{Y}_{j,\bm\delta}$ comes from~\eqref{eq:constraints_slack} and~\eqref{eq:MPC_iob} is the constraint that takes into consideration the IOB, from~\eqref{eq:constraints_u2_slack}. The tightened constraints are computed as explained in the previous section, for all the subjects, and the resulting~$N_c$ are reported in Table~\ref{tb:mpc_settings}.
The prediction horizon is set to~$N_p=12$ for all subjects, to reach 60 minutes of predictions. 

The cost function $V_N(\hat{x},u; \Theta, \mathcal{D})$ to be minimized is designed by summing several components, namely: 
\begin{equation}
    V_N(\hat{x},u; \Theta, \mathcal{D})=V_{N_c}+V_{N_p}+V_s+\lambda V_P+V_{\delta}+V_u .
\end{equation}

Specifically, the stage costs~$V_{N_c}$, along the control horizon~$N_c-1$, and~$V_{N_P}$, along the prediction horizon~$N_p-1$ starting from~$N_c$, are:
\begin{subequations}
    \begin{equation} 
V_{N_c}=\sum_{j=0}^{N_c-1}\|\hat{y}(j|k)-y_a\|^2_Q + \|u_2(j)-u_\mathrm{ref}\|_R^2  ,   
\end{equation}
\begin{equation}
V_{N_p}=\sum_{j={N_c}}^{N_p-1} \|\hat{y}(j|k)-y_a\|_Q^2 .
\end{equation}
\end{subequations}

where the insulin target $u_\mathrm{ref}$ is the constant basal insulin dose delivered by the UVA/Padova simulator for the default continuous therapy of the selected virtual patient. 
For the implementation of the MPC in a zone control fashion is required the presence of the setpoint $y_a$, which is an auxiliary optimization variable and it has to be within 70 and 140~\si{\milli\gram / \deci\liter}.
Two slack variables $\delta_\mathrm{hypo}$ and $\delta_\mathrm{hyper}$ are added to this interval for $y_a$, which are other optimization variables used to increase the range in the constraints when necessary. Therefore, an additional stationary cost $V_s$ has to be considered. 
In $V_s$, the slack variables are weighted by some constants, that are set to be~$p_\mathrm{hypo} > p_\mathrm{hyper}$ to reflect the higher danger of hypoglycemia compared to hyperglycemia~\cite{abuin2020artificial}.
\begin{equation}
	V_s=p_\mathrm{hyper}\delta_\mathrm{hyper}^2 + p_\mathrm{hypo}\delta_\mathrm{hypo}^2  .
\end{equation}

The terminal cost $V_P$ penalises the difference between the last state $\hat{x}(N_p|k)$ and the reference state ($x_\mathrm{ref}$, which contains the set point $y_a$, no meals and $u_\mathrm{ref}$). 
\begin{equation} \label{eq:terminal_cost}
    V_P = \|\hat x(N_p|k)-x_\mathrm{ref}\|_P^2 ,
\end{equation}
where $P$ is the solution to the Riccati equation, given the LQR control gain $K$ for the linearized system around the reference point (in Section~\ref{sec:terminal_ingredients}).
The terminal cost is normally used to ensure the stability of the MPC and in this case, it is weighted by a factor~$\lambda>0$, as no terminal constraint is taken into account.

Other slack optimisation variables are considered in the glycemic constraints~\eqref{eq:constraints_slack}. Thus, for the same reason as before, the cost~$V_{\delta}$ must be included to penalise them by considering two weights, where $p_\mathrm{min}>p_\mathrm{max}$,
\begin{equation}
         V_{\delta}=\sum_{j=1}^{N_P} \| \delta_\mathrm{min}(j)\|^2_{p_\mathrm{min}}  +  \| \delta_\mathrm{max}(j)\|^2_{p_\mathrm{max}} .
\end{equation}

The last component is the cost $V_u$, to penalise additional slack variables $\delta_u$ which are included in the control action constraints~\eqref{eq:MPC_iob}, 
\begin{equation}
        V_u=\sum_{j=1}^{N_P}  \| \delta_u(j)\|^2_{p_u} .
\end{equation}

Furthermore, many combinations of weights were tested and the following were selected: $R=10$, $p_\mathrm{hypo}=1 \cdot 10^7$, $p_\mathrm{hyper}=1 \cdot 10^6$, $p_\mathrm{min}=1 \cdot 10^7$, $p_\mathrm{max}=1 \cdot 10^6$,  $p_u=1 \cdot 10^7$ and $\lambda=10$. The values of $Q$ are shown in Table~\ref{tb:mpc_settings} and note that in the cases where $R$ is greater than $Q$, this means that a more conservative controller is applied.

%% file: 4_results.tex
\section{Results}
\label{sec:results}
\begin{figure*}
\begin{center}
\includegraphics[width=\linewidth]{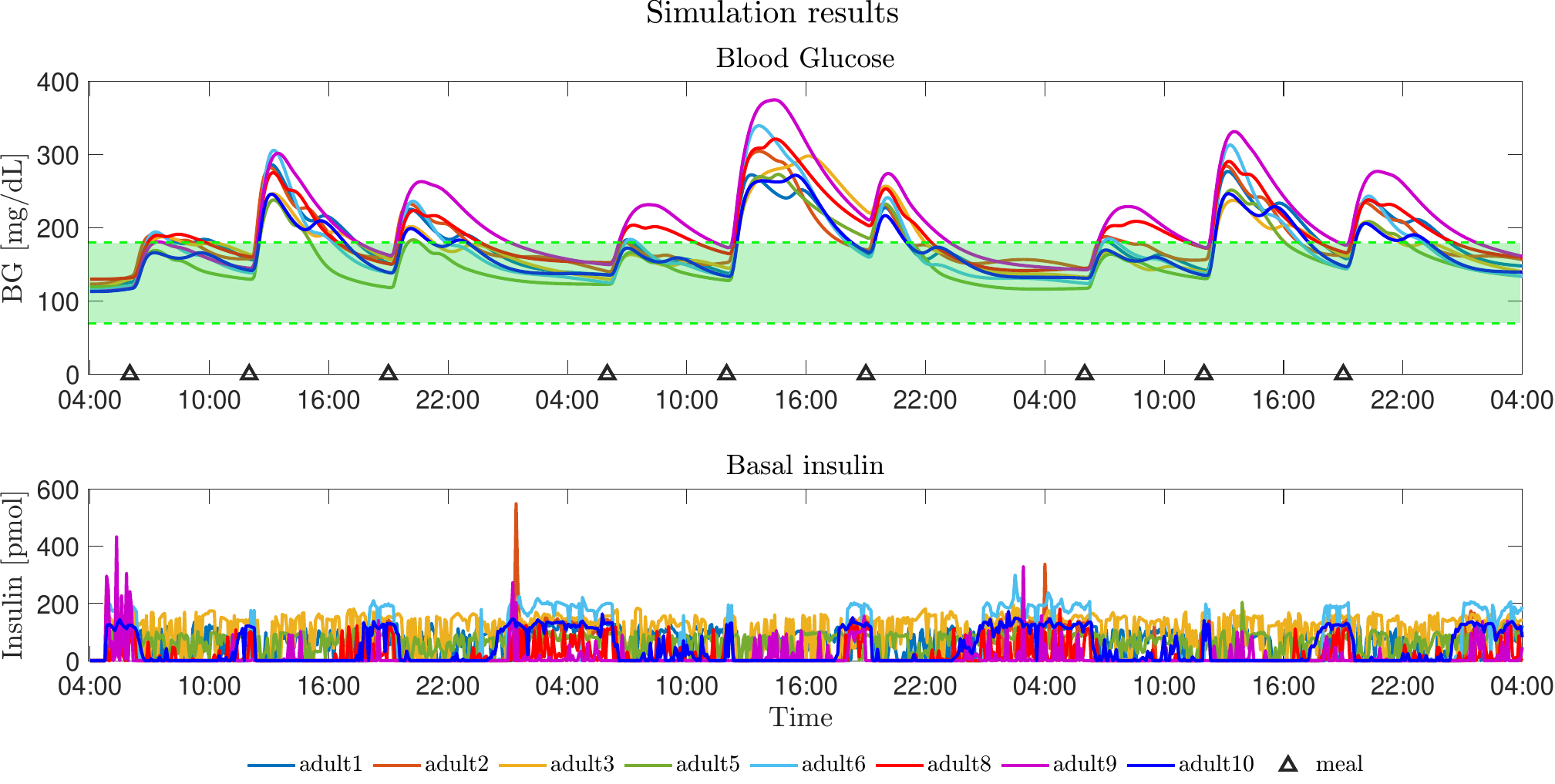}
\caption{The upper plot displays BG trends for all patients. The green zone represents the safe range and the black triangles depict meals. The lower plot shows basal insulin injections computed by the proposed MPC.}
\label{fig:bg_ins_iob}
\end{center}
\end{figure*}

The proposed MPC was tested on the virtual adult patients of the UVA/Padova simulator, with customized controllers for each subject.
Three days were simulated, each day consisting of the following three meals: \SI{40}{\gram} of carbohydrates at 06:00 am, \SI{100}{\gram} at 12:00 pm and \SI{60}{\gram} at 07:00 pm, with a duration of~\SI{15}{\minute}. The postprandial boluses were computed by the simulator and injected~\SI{20}{\minute} after the start of the meal. All the devices have the same noise setting as in the data acquisition stage.

The results of the simulations for all patients are shown in Figure~\ref{fig:bg_ins_iob}. The top graph displays the BG trends caused by the insulin injections depicted in the bottom graph, which vary based on the patient model.
The primary objective is to reduce the frequency and the severity of hypoglycemic events, which are very dangerous in the short term, and it can be seen that such a result is achieved.

\begin{figure}
    \centering
    \includegraphics[width=0.9\linewidth]{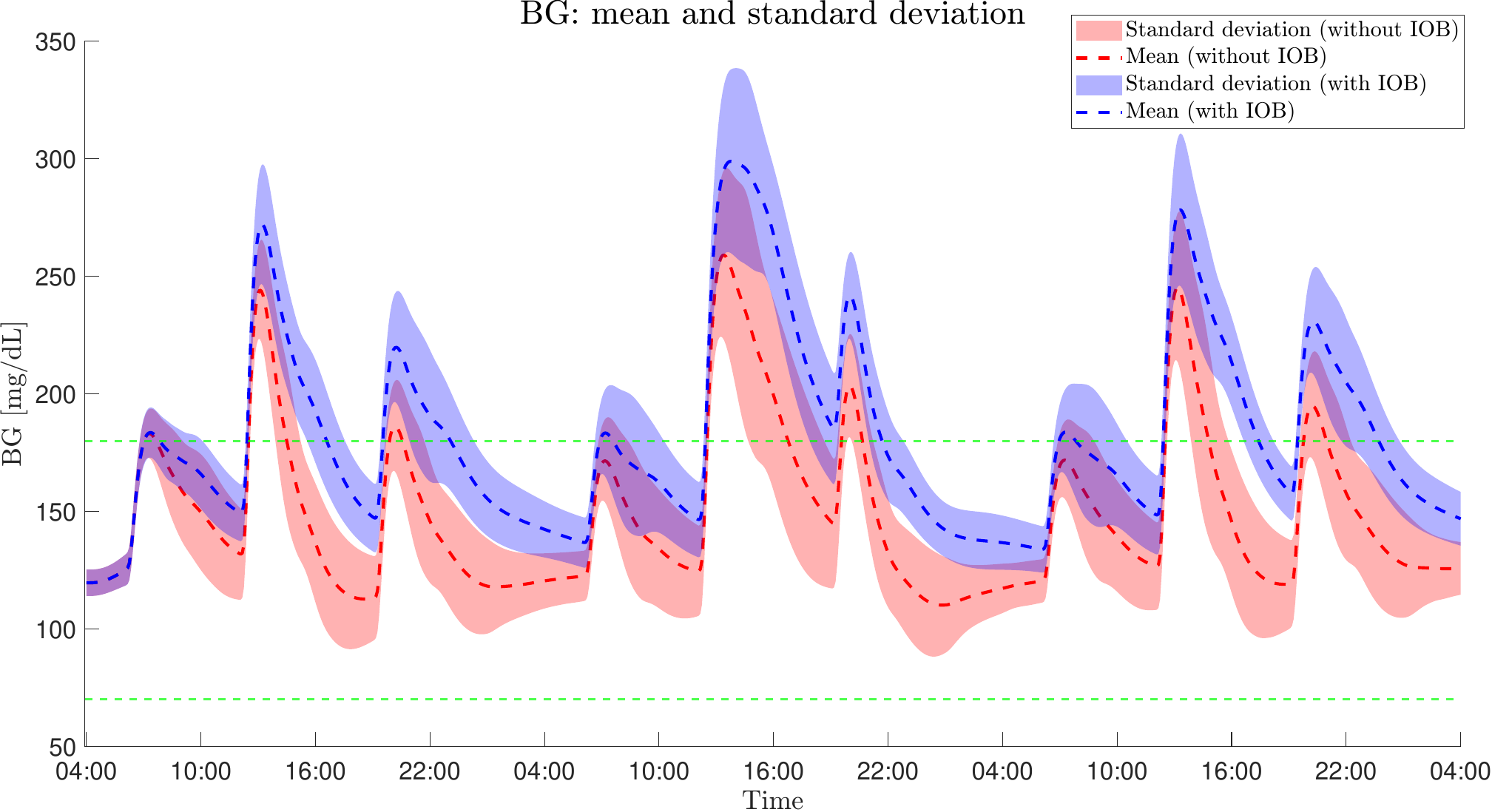}
    \caption{Comparison of the BG values: the simulations performed with IOB constraints are represented in blue, and without them are in red.}
    \label{fig:BG_mean_std_comparison_IOB}
\end{figure}

The results obtained in this work are compared to the ones reported in~\cite{ifac}, whose simulations are conducted without the IOB constraints. 
In particular, the comparison of the mean and standard deviation of the BG values for the two cases is shown in Figure~\ref{fig:BG_mean_std_comparison_IOB}, where the blue dotted line represents the mean values of the cases with the IOB constraints and the blue area displays their standard deviations, while in red the cases without the IOB.
This indicates that including the IOB in the constraints means to have a more conservative controller, since the BG level is higher in the cases with the IOB safety constraints. It is also confirmed looking at the BG and basal insulin average values reported in Table~\ref{tb:slack_iob} (i.e. mean and standard deviation), where the average BG values of the simulations with IOB constraints are higher than in the ones without them, because of less insulin amount.
This is due to the fact that the controller does not manage post-prandial boluses. As a result, when the BG value is high at mealtime, the bolus amount tends to be higher (computed by the simulator). This leads to higher IOB, which in turn limits the basal corrections.

\begin{figure}
    \centering
    \includegraphics[width=\linewidth]{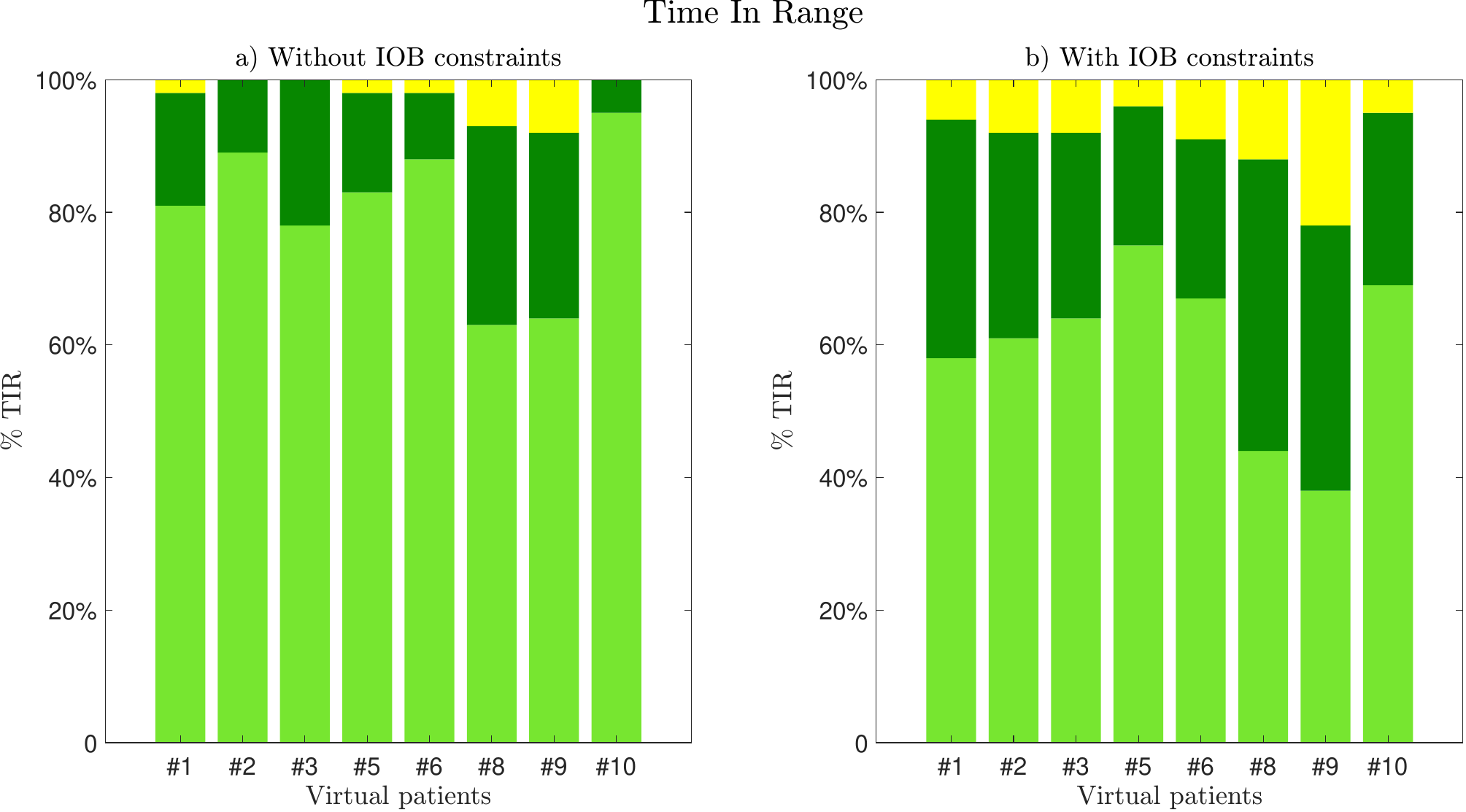}
    \includegraphics[width=0.65\linewidth]{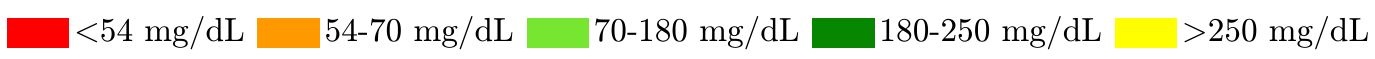}
    \caption{TIR results of the simulations performed with (graph on the right) and without the IOB constraints (graph on the left). Each bar represents a specific subject.}
    \label{fig:TIR_slack_iob}
\end{figure}

Another important tool for assessing AP performance is the Time In Range (TIR). This shows the percentage of time a patient spends in each specific BG range. In particular, as required by the American Diabetes Association, the TIR targets are as follows:~$<5\%$ of time with BG higher than 250~\si{\milli\gram / \deci\liter}, $<25\%$ between 180-250~\si{\milli\gram / \deci\liter}, $>70\%$ between 70-180~\si{\milli\gram / \deci\liter}, $<4\%$ between 55-70~\si{\milli\gram / \deci\liter} and $<1\%$ for BG lower than 55~\si{\milli\gram / \deci\liter}.
The proposed controller ensures that the requirements for the hypoglycemic ranges are always met, which is the main objective. However, the controller permits the subjects to stay a little longer in the two hyperglycemic ranges, which also means that they stay within the 70-180~\si{\milli\gram / \deci\liter} range for less than 70\% of the simulation time.
The results are displayed in Figure~\ref{fig:TIR_slack_iob}, where, for each subject, the graph on the left is for the cases without the IOB, while the one on the right is for the cases with the IOB constraints.

The conservative results are a consequence of the linear function used to estimate the IOB. In fact, as shown in Figure~\ref{fig:IOB_estimation_1}, the IOB is always overestimated. This can be improved searching for a polynomial or exponential function. 

An additional useful tool can be the Glycemia Risk Index (GRI) \cite{klonoff2022glycemia}, which is a quantitative measure designed to provide a comprehensive evaluation of an individual's susceptibility to hypoglycemia or hyperglycemia. It is obtained from 
\begin{equation} \nonumber
    GRI = \Big(3.0\big(p_1+0.8 p_2\big)\Big)+\Big(1.6\big(p_4+0.5 p_3\big)\Big) ,
\end{equation}
where the first part is the hypoglycemic component and the second is the hyperglycemic one. It considers the same percentages of the TIR, where $p_1$ is the percentage of time in which the subject's BG is less than 54 mg/dL, $p_2$ for the BG between 54 and 70 mg/dL, $p_3$ for the BG between 180 and 250 mg/dL and $p_4$ for BG higher than 250 mg/dL. 
The GRI can be displayed graphically on a grid with the hypoglycemia component on the horizontal axis and the hyperglycemia component on the vertical axis. Diagonal lines divide the graph into five zones (quintiles) based on overall glycemia quality, from best ($0^\mathrm{th}$-$20^\mathrm{th}$ percentile) to worst ($81^\mathrm{st}$-$100^\mathrm{th}$ percentile).

\begin{figure}
    \centering
    \includegraphics[width=\textwidth]{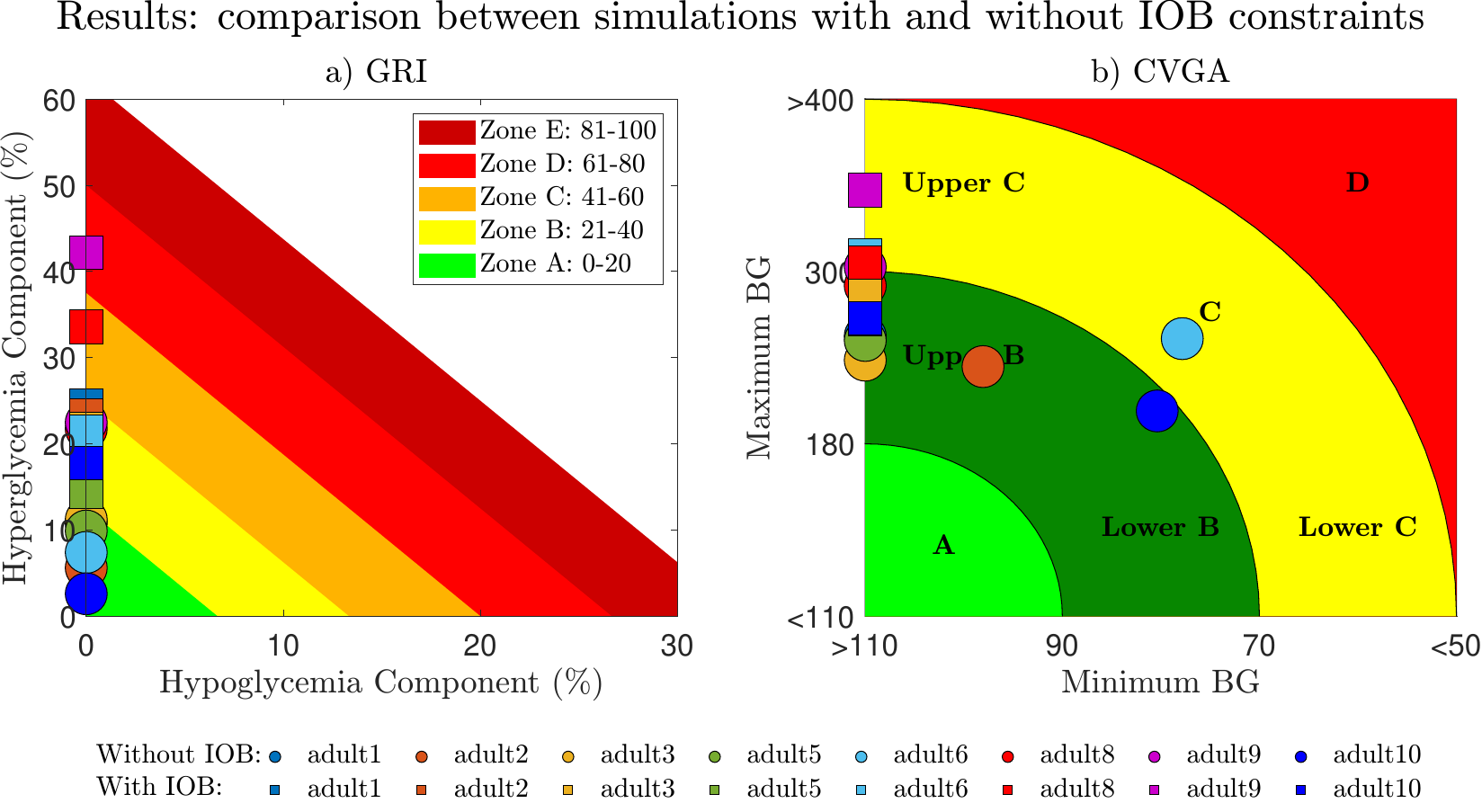}
    \caption{Results of the simulations performed with the IOB constraints (squares), compared to the ones without them (dots): a) GRI and b) CVGA.}
    \label{fig:GRI_CVGA_withIOB}
\end{figure}

The GRI values are reported in Table~\ref{tb:slack_iob}, while the two components are represented in Figure~\ref{fig:GRI_CVGA_withIOB}a to understand which is the higher one, where each dot on the graph describes a specific subject in the cases without the IOB, and the squares are for ones with the IOB constraints. 
In the cases without the IOB, Adult 8 and Adult 9 are in Zone B, while the others are in safe Zone A.
While in the cases with IOB, due to the higher BG values, Adult 8 is in Zone C and Adult 9 is in Zone D, while the others are in Zone B.
All the subjects lay on the $y$-axis, this is because our controller is designed to avoid hypoglycemia, which is why the higher risk component is the hyperglycemic one.

Up to now, the average results are evaluated, but to have a more complete analysis, also the Control-Variability Grid Analysis (CVGA) can be assessed.
The CVGA \cite{magni} is a graphical representation that provides both visual and numerical information about the quality of glycemic control. 
In Figure~\ref{fig:GRI_CVGA_withIOB}b, each dot on the graph describes a specific subject in the cases without IOB, while the squares are for those with IOB restrictions, with the minimum BG value as the $x$-coordinate and the maximum BG value as the $y$-coordinate. 
Considering the simulations with the IOB, these worst cases are all in the safe zones (except for Adult 6, Adult 8 and Adult 9, who are in Zone C). 

\begin{table}[]
\centering
\caption{Comparison of the BG mean and standard deviation (in~\si{\milli\gram / \deci\liter}), of the basal insulin $u_2$ mean and standard deviation (in~\si{\pico\mol}), and GRI for the controller with (on the right) and without (on the left) the IOB constraints.}
\label{tb:slack_iob}
\begin{tabular}{|c|c|c|c|c|c|c|}
\hline
\multicolumn{1}{|c|}{\multirow{2}{*}{Adult}} & \multicolumn{3}{c|}{Without IOB}   & \multicolumn{3}{c|}{With IOB}           \\ \cline{2-7} 
\multicolumn{1}{|c|}{}                       & \multicolumn{1}{c|}{BG mean $\pm$ std} & \multicolumn{1}{c|}{$u_2$ mean $\pm$ std} &  \multicolumn{1}{c|}{GRI} & \multicolumn{1}{c|}{BG mean $\pm$ std} & \multicolumn{1}{c|}{$u_2$ mean $\pm$ std}& \multicolumn{1}{c|}{GRI} \\ \hline
\#1  & $158.18 \pm 32.31$  & $95.21 \pm 45.70$    & 17.29  & $180.33 \pm 38.48 $   &   $71.76 \pm 47.92$   &  39.19  \\
\#2  & $138.72 \pm 29.91$  & $113.17 \pm 57.17 $  & 8.97   & $183.90 \pm 38.77$  & $34.64 \pm 60.93$    &  37.18  \\
\#3  & $158.30 \pm 31.79$  & $129.57 \pm  47.69$  & 17.94  & $178.35 \pm  40.34$  &  $107.53 \pm 61.75$  &  34.77  \\
\#5  & $149.51 \pm 35.80$  & $76.23 \pm 32.81$    & 15.82  & $159.97 \pm 39.44$   & $67.03 \pm 32.79$    & 23.12   \\
\#6  & $127.47 \pm 42.09$  & $182.95 \pm 42.63$   & 11.84  & $175.33 \pm 49.99$   & $51.78 \pm 83.83$   &   34.22 \\
\#8  & $171.06 \pm 41.89$  & $52.96 \pm 66.67$    & 34.77  & $193.77 \pm  53.73$  &   $17.90 \pm 36.08$   &  53.73  \\
\#9  & $168.86 \pm 50.44$  & $48.39 \pm 68.67$    & 35.98  & $208.87 \pm 56.09$   &   $10.74 \pm 38.51$   & 67.51   \\
\#10 & $120.11 \pm 28.79$  & $119.40 \pm 27.15$   & 4.16   & $170.25 \pm 35.93$   &  $43.38 \pm 56.11$    &  28.39   \\ \hline
\end{tabular}
\end{table}

\subsection{Simulations with insulin sensitivity variations}
\begin{figure}
    \centering
    \includegraphics[width=\linewidth]{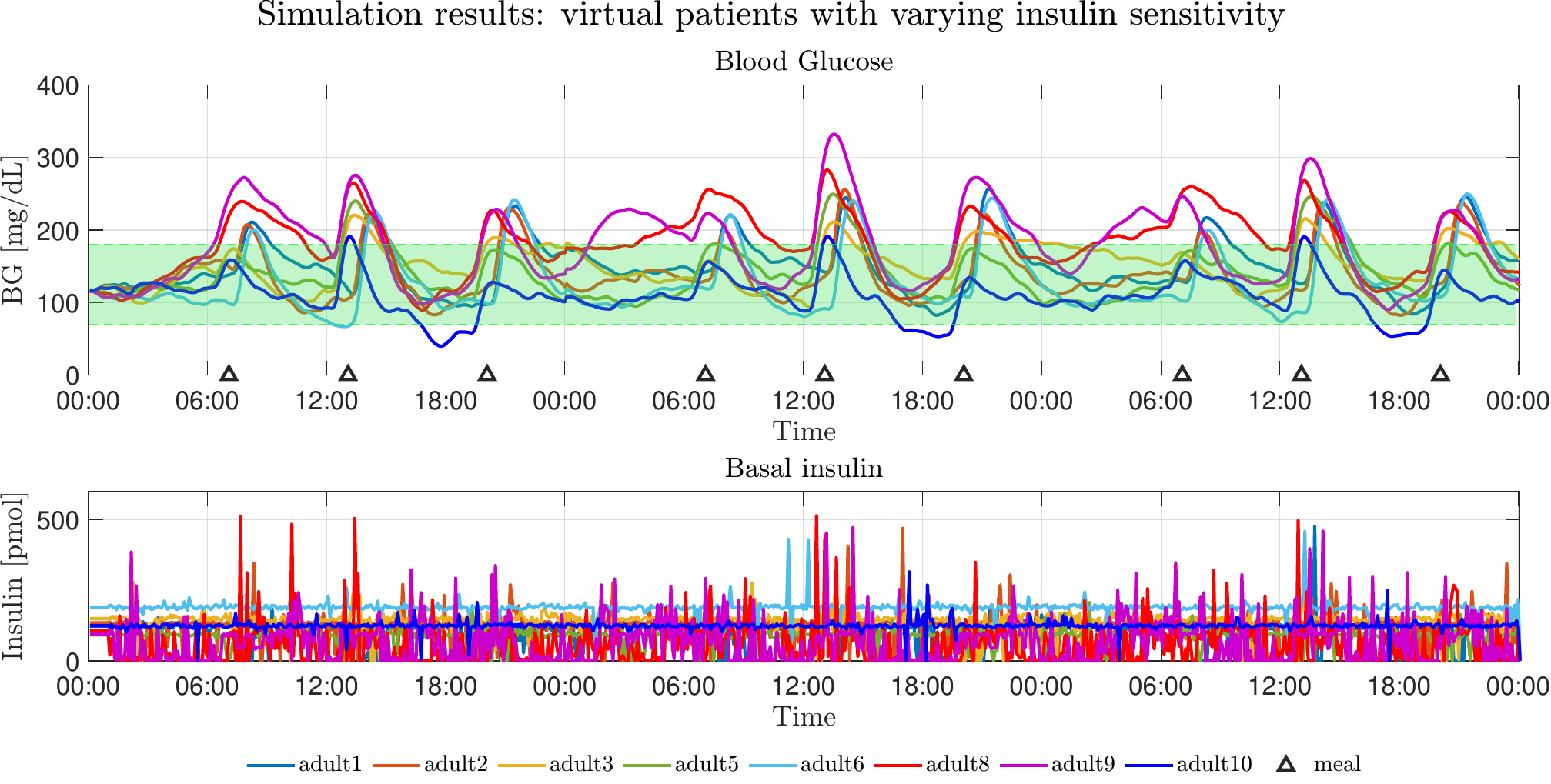}
    \caption{Proposed controller applied to the virtual patients with varying insulin sensitivity. Upper plot: BG trends of all patients, with the green zone for the safe range and the black triangles for the meal times. Lower plot: basal insulin injections.}
    \label{fig:bg_ins_slack_iob_ins}
\end{figure}

\begin{figure}
    \centering
    \includegraphics[width=\linewidth]{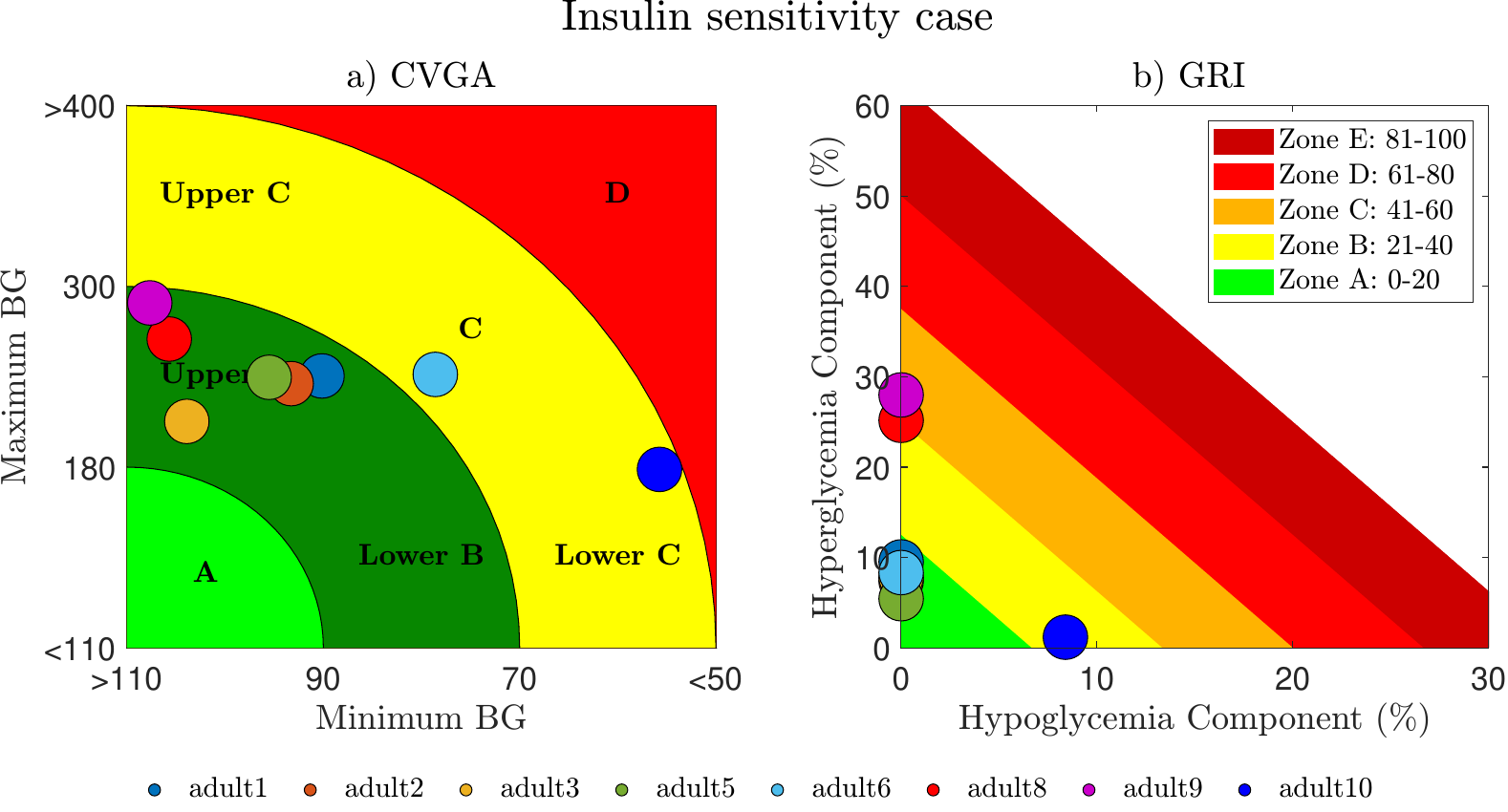}
    \caption{Results of the simulations performed with the IOB constraints, applied to the varying insulin sensitivity case: a) CVGA and b) GRI.}
    \label{fig:cvga_gri_slack_iob_ins}
\end{figure}

\begin{figure}
    \centering
    \includegraphics[width=0.6\linewidth]{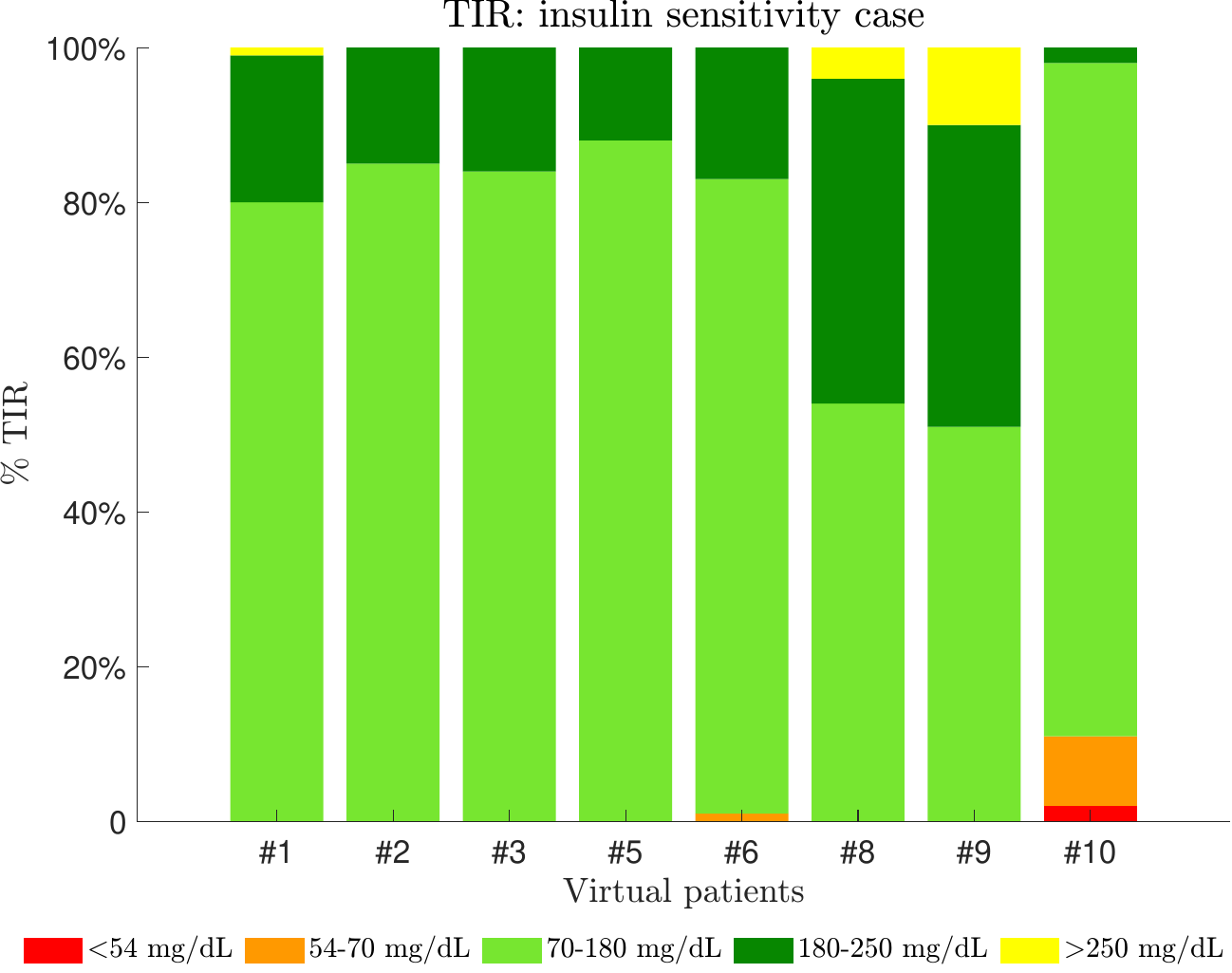}
    \caption{TIR results of the simulations performed with the IOB constraints, applied to the varying insulin sensitivity case.}
    \label{fig:tir_slack_iob_ins}
\end{figure}

In this section, the proposed CHoKI-based MPC with the IOB constraints is tested on the same virtual patients, but with variations in the insulin sensitivity.
Insulin sensitivity refers to how responsive the cells are to insulin and this can vary in the subject during the day~\cite{visentin2015circadian}.
The simulations are performed with the same setting as in the previous cases and the results are presented in Figure~\ref{fig:bg_ins_slack_iob_ins}, where the upper part shows the BG trends that are obtained thanks to the insulin injections displayed in the lower graph. 
To better evaluate the performances of the proposed controller in managing the variations in the insulin sensitivity, the CVGA, GRI and TIR results are evaluated as well (see  Figure~\ref{fig:cvga_gri_slack_iob_ins}a, Figure~\ref{fig:cvga_gri_slack_iob_ins}b and Figure~\ref{fig:tir_slack_iob_ins}, respectively). 
The results are quite promising, but the variations in insulin sensitivity affect the ability of the CHoKI learning method. This is visible for example in Adult 10, who undergoes three hypoglycemic events. 

The CHoKI technique estimates the parameters $\mathcal{L}$ and $\mathcal{P}$ once and offline. To better manage the insulin variability a possible solution could be to use an adaptive CHoKI, which can update $\mathcal{L}$ and $\mathcal{P}$ values depending on the blood glucose and insulin situation.

%% file: 5_conclusion.tex
\section{Conclusion}
\label{sec:conclusion}
A new MPC algorithm based on the CHoKI learning method was proposed to be used in the AP for managing basal insulin in T1D patients, including IOB estimation to limit the amount of basal insulin injections.
The whole system was tested on the virtual patients of the UVA/Padova simulator.
The proposed controller aims to drive and maintain the BG level inside the euglycemic range most of the time, trying to avoid the more dangerous hypoglycemic events.
The obtained results seem promising, since the estimation of the IOB in the MPC helps in achieving such an aim. 
To decrease the level of the hyperglycemic events, the IOB estimation could be improved. In particular, a polynomial or exponential curve can be tried instead of the linear weights employed in this case.

The proposed controller was also tested on virtual patients with variability in insulin sensitivity. To improve these results, the next step for future works could be to identify multi-models, dividing the day into some intervals (such as breakfast, lunch and dinner) and trying to learn different behaviour, with the aim of controlling patients more accurately, thanks to the inclusion of the analysis of insulin sensitivity variations during the day.

\section*{Acknowledgement} 
This work was funded by the National Plan for NRRP Complementary Investments (PNC, established with the decree-law 6 May 2021, n. 59, converted by law n. 101 of 2021) in the call for the funding of research initiatives for technologies and innovative trajectories in the health and care sectors (Directorial Decree n. 931 of 06-06-2022) - project n. PNC0000003 - AdvaNced Technologies for Human-centrEd Medicine  (project acronym: ANTHEM). This work reflects only the authors' views and opinions, neither the Ministry for University and Research nor the European Commission can be considered responsible for them.